\documentclass[final,5p,times,twocolumn]{elsarticle}
\usepackage{graphicx}
\usepackage{amssymb}
\journal{Physical Letters B}
\begin{document}
\begin{frontmatter}
\title{Consistent threshold $\pi^0$ electro-production at
  $Q^2=0.05$, 0.10, and
  $0.15\,\mathrm{GeV}^2/c^2$}
\author[kph]{H.~Merkel\corref{cor1}}
\ead{merkel@kph.uni-mainz.de}
\cortext[cor1]{Corresponding author}
\author[kph]{P.~Achenbach}              \author[kph]{C.~Ayerbe~Gayoso}
\author[kph]{M.~Ases~Antelo}            \author[kph]{D.~Baumann}
\author[mit]{A.\,M.~Bernstein}          \author[kph]{R.\,B\"ohm}
\author[zagreb]{D.~Bosnar}              \author[kph]{M.~Ding}
\author[kph]{M.\,O.~Distler}            \author[kph]{L.~Doria}
\author[kph]{J.\,Garcia~Llongo}         \author[jlab]{D.\,W.~Higinbotham}
\author[kph]{G.~Jover~Ma{\~n}as}        \author[zagreb]{M.~Makek}
\author[kph]{U.~M\"uller}               \author[kph]{R.~Neuhausen}
\author[kph]{L.~Nungesser}              \author[kph]{R.~P\'erez~Benito}
\author[kph]{J.\,Pochodzalla}           \author[kph]{M.\,Seimetz}
\author[stefan,ljubljana]{S.~\v{S}irca} \author[mit]{S.~Stave}
\author[kph]{Th.~Walcher}               \author[kph]{M.~Weis}
\address[kph]{Institut f\"ur Kernphysik, 
              Johannes Gutenberg-Universit\"at Mainz, D-55099~Mainz, Germany.}
\address[mit]{Laboratory for Nuclear Science, 
              Massachusetts Institute of Technology, Cambridge, MA~02139, USA.}
\address[zagreb]{Department of Physics, 
              University of Zagreb, HR-10002 Zagreb, Croatia.}
\address[jlab]{Thomas Jefferson National Accelerator Facility, 
              Newport News, VA 23606, USA.}
\address[stefan]{Jo\v{z}ef Stefan Institute, SI-1001 Ljubljana, Slovenia.}
\address[ljubljana]{Department of Physics, 
              University of Ljubljana, SI-1000 Ljubljana, Slovenia.}
\begin{abstract}
  New, accurate data are presented on the near threshold
  $p(e,e'p)\pi^0$ reaction in the range of four-momentum transfers
  between $Q^2$=0.05 and 0.15\,$\mathrm{GeV^2/c^2}$. The data were
  taken with the spectrometer setup of the A1 Collaboration at the
  Mainz Microtron MAMI. The complete center of mass solid angle was
  covered up to a center of mass energy of $4\,\mathrm{MeV}$ above
  threshold. These results supersede the previous analysis based on
  three separate experiments, and are compared with calculations in
  Heavy Baryon Chiral Perturbation Theory and with phenomenological
  models.
\end{abstract}
\end{frontmatter}

\section{Introduction}

Threshold electromagnetic pion production is a fundamental process
since the pion is a Nambu-Goldstone boson due to the spontaneously
broken chiral symmetry of QCD \cite{book}. Calculations at low energies
which are good approximations to QCD are carried out by an effective
field theory called Chiral Perturbation Theory (ChPT)
\cite{Weinberg:1978kz, Gasser:1983yg, Gasser:1983kx}, and are
generally in good agreement with experiment. The systematic
application of Heavy Baryon Chiral Perturbation Theory (HBChPT)
\cite{Bernard:2006gx,Bernard:2007zu} has been generally successful in
describing $\pi-N$ scattering and electromagnetic pion production from
the nucleon \cite{Bernstein:2009dc}.

In recent years there has been a considerable experimental effort to
test this theoretical approach. The latest published and most accurate
of a series of experiments of photo-pion production experiments which
was performed at Mainz \cite{Schmidt:2001vg} was in good agreement with
the HBChPT \cite{Bernard:1994gm}\footnote{A more recent accurate
  measurement at Mainz has been performed and is in the final stages
  of data analysis \cite{hornidge}}. These
experimental tests can also be extended to four-momentum transfer $Q^2
> 0$ via pion electro-production. This adds another small scale whose
convergence properties are not presently known.

\begin{figure}
\centerline{\includegraphics[width=0.8\columnwidth]{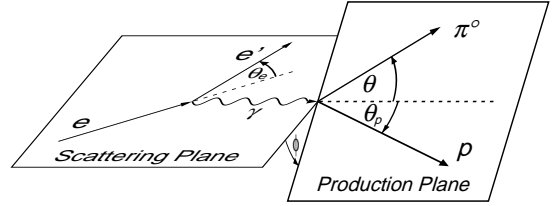}}
\caption{Definition of angles.}
\label{figplanes}
\end{figure}

The first threshold neutral pion electro-production experiments were
performed at NIKHEF \cite{Welch:1992ex,vandenBrink:1997cs} and
MAMI \cite{Distler:1998ae} at $Q^2 = 0.10\,\mathrm{GeV^2/c^2} \approx 5
m_{\pi}^{2}$ and the results were in reasonable agreement with the
calculations \cite{BKM96a}. A second measurement at
$Q^2=0.05\,\mathrm{GeV^2/c^2}$ at Mainz \cite{Merkel:2001qg} indicated
a surprisingly rapid $Q^{2}$ variation which was not in agreement with
the calculations \cite{BKM96a}. However, this conclusion was
problematic for three reasons. First, the observed variation was based
on three independent experiments \cite{Schmidt:2001vg, Distler:1998ae,
  Merkel:2001qg}. Second, the data analysis programs that were used
have been subsequently revised. Finally, a subsequent, independent
experiment performed at Mainz at $Q^2=0.05\,\mathrm{GeV^2/c^2}$ was
extended to higher energies above threshold ($\Delta W <
40\,\mathrm{MeV}$ at a center of mass pion production angle of
$90^{0}$) \cite{Weis:2007kf}. This most recent experiment disagreed
with the previous values \cite{Merkel:2001qg} in the near threshold
region, and was not in agreement with the calculations based on
HBChPT \cite{BKM96a}. It was also not in agreement with the
phenomenological MAID model \cite{DHKT99}, but was in generally good
agreement with the DMT model which uses a chiral
Lagrangian \cite{PhysRevLett.83.4494,PhysRevC.64.032201}.

To resolve the experimental uncertainties of the previous experiments,
a new measurement of the variation of the $\pi^0$ electroproduction
cross section was performed in the near threshold region ($\Delta W
\leq 5\,\mathrm{MeV}$) for $Q^2 = 0.05, 0.10,
0.15\,\mathrm{GeV}^2/c^2$. These new results are the subject of this
article and will be compared to model
\cite{DHKT99,PhysRevLett.83.4494,PhysRevC.64.032201} and to HBChPT
\cite{BKM96a} calculations. We point out that since several of the low
energy constants of HBChPT have been fitted to the older data with
significant experimental errors, these calculations should not be
considered as prediction and will have to be re-adjusted to take this
into account.  For this reason these calculations are not projected to
$Q^{2} > 0.10\,\mathrm{GeV}^2/c^{2}$.

\section{Formalism}

In the one photon exchange approximation, the electro-production cross
section of pseudo-scalar mesons can be written as (see
e.g. \cite{DT92})
\newlength\nmln\settoheight\nmln{$\sqrt{2\epsilon(1+\epsilon)}$}
\begin{eqnarray}
  \frac{d^5\sigma(\theta,\phi)}{dE_e d^2\Omega_e d^2\Omega} &=& 
  \Gamma \left(\rule{0mm}{\nmln}
                \sigma_{\mathrm{T}}(\theta) 
     + \epsilon~\sigma_{\mathrm{L}}(\theta)\right.\nonumber\\ 
   &+& \epsilon~\sigma_{\mathrm{TT}}(\theta)\cos2\phi\nonumber\\[1mm] 
  &+& \left.\!\!\sqrt{2\epsilon(1+\epsilon)}~\sigma_{\mathrm{LT}}(\theta)\cos\phi\right),
  \label{equ:cross}
\end{eqnarray}
with the virtual photon flux
\begin{equation} 
  \Gamma = \frac{\alpha}{2\pi^2}\frac{E'}{E} 
           \frac{k_\gamma}{(-q^2)}\frac{1}{1-\epsilon}
\end{equation}
and the transverse photon polarization
\begin{equation}
  \epsilon =
  \left(1-\frac{2\left(\omega^2-q^2\right)}{q^2}
    \tan^2\frac{\theta_e}{2}\right)^{-1}.
\end{equation}
The energy of the initial and scattered electron in the laboratory
frame is given by $E$ and $E'$, the photon equivalent energy is
defined as $k_\gamma = (W^2-m_p^2)/(2m_p)$. The photon four-momentum
transfer is $q^2= - Q^2= \omega^2-\vec{q}^2$, with the photon
laboratory energy and momentum $\omega$ and
$\vec{q}$. Figure~\ref{figplanes} shows the definition of the pion
production angles $\theta$ and $\phi$, which are given in the
proton-photon center of mass system (CMS) in the following. In
addition, the cross sections also depend on the center of mass energy
$W$ (or $\Delta W = W - m_{\pi^0}-m_p$).

Without variation of the photon polarization $\epsilon$,
$\sigma_{\mathrm{T}}(\theta)$ and $\sigma_{\mathrm{L}}(\theta)$ cannot
be separated and only the unseparated cross section
\begin{equation}
  \sigma_0(\theta) = \sigma_{\mathrm{T}}(\theta) + \epsilon \sigma_{\mathrm{L}}(\theta)
\end{equation}
can be extracted.

\begin{figure}
\includegraphics[width=\columnwidth]{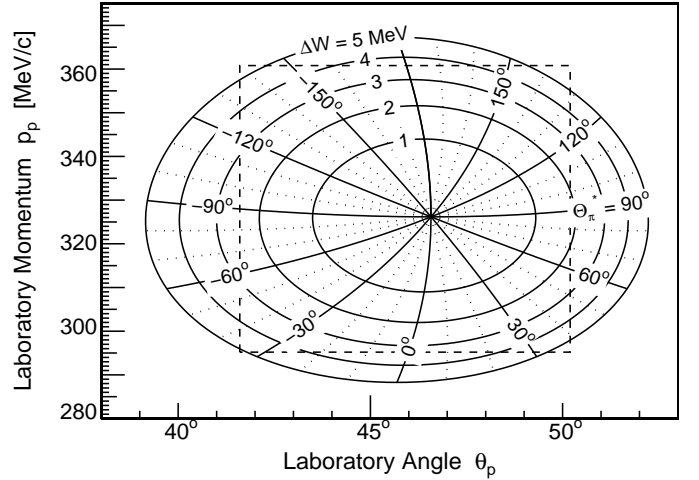}
\caption{Acceptance of the proton for $Q^2=0.10\,\mathrm{GeV^2}/c^2$.
  Lines of constant center of mass energy $\Delta W$ and constant
  center of mass production angle $\theta_\pi$ are drawn.  The dashed
  line shows the acceptance of spectrometer A (including energy loss
  corrections).}
\label{figacceptance}
\end{figure}

\section{Experiment}

The experiment was performed at the spectrometer setup of the A1
collaboration at MAMI (see Ref.~\cite{Blo98} for a detailed
description of the setup). The MAMI accelerator delivered an
unpolarized electron beam with an energy of 855\,MeV and a beam
current of up to 9\,$\mu$A to an oblong liquid Hydrogen target cell
with a width of 1\,cm and a length of 5\,cm. The beam was rastered
across the target cell to avoid local boiling of the liquid
hydrogen. A luminosity of $L=1.2\cdot
10^{37}\,\mathrm{s}^{-1}\mathrm{cm}^{-2}$ was achieved.

\begin{table}
\caption{Kinematic setups. The beam energy was
  $E=855\,\mathrm{MeV}$ for all three setups.}
\label{kinematics}
\begin{tabular*}{\columnwidth}{@{\extracolsep{\fill}}c@{}cc@{~~}cc@{~~}c}
  \hline
  \multicolumn{2}{c}{Photon} &
  \multicolumn{2}{c}{Proton (A)} &
  \multicolumn{2}{c}{Electron (B)} \\
  $Q^2$ & $\epsilon$ & $\theta_p$ & $p_p$ & $\theta_e$ & $E'$ \\
$(\mathrm{GeV}^2/c^2)$ 
&&& $(\mathrm{MeV}/c)$ && $(\mathrm{MeV})$ \\
\hline
0.05 & 0.932 & 45.7$^\circ$  & 235.8 & 16.8$^\circ$ & 683.5 \\
0.10 & 0.882 & 45.9$^\circ$  & 320.0 & 24.4$^\circ$ & 652.9 \\
0.15 & 0.829 & 45.5$^\circ$  & 387.9 & 30.6$^\circ$ & 630.0 \\
\hline
\end{tabular*}
\end{table}

The scattered electron was detected by Spectrometer B with an angular
acceptance of 5.6\,msr and a momentum acceptance of $\Delta p/p =
15\%$. The recoil proton was detected by Spectrometer A with an
angular acceptance of 21\,msr and a momentum acceptance of $\Delta p/p
= 20\%$. Three different setups were chosen corresponding to three
values of the four-momentum transfer $Q^2$, table \ref{kinematics}
summarizes the kinematic setups. Due to the relativistic boost of the
center of mass system, nearly the complete solid angle fits within the
acceptance of spectrometer A in a single
setup. Figure~\ref{figacceptance} shows the acceptance in the
laboratory system with lines of constant CMS energy and constant CMS
angle for the intermediate value of the four-momentum transfer
$Q^2=0.10\,\mathrm{GeV}^2/c^2$. As can be seen, complete acceptance can
be reached up to nearly $\Delta W=4\,\mathrm{MeV}$.

At the focal plane of the spectrometers, vertical drift chambers were
used for the position and angular reconstruction of the particle
trajectories. Two layers of scintillators were used for trigger
purpose and coincidence time measurement. In addition, threshold gas
\v{C}erenkov detectors were used to verify that there are no sizable
contributions of pions in the electron or proton arm. However, they
were not used in the final analysis to reduce systematic errors due to
the determination of the inhomogeneous efficiency of these
detectors. Both spectrometers reach a momentum resolution (FWHM) of
$\delta p/p < 10^{-4}$ and an angular resolution of better than
0.2\,mrad.

\begin{figure}
\includegraphics[width=\columnwidth]{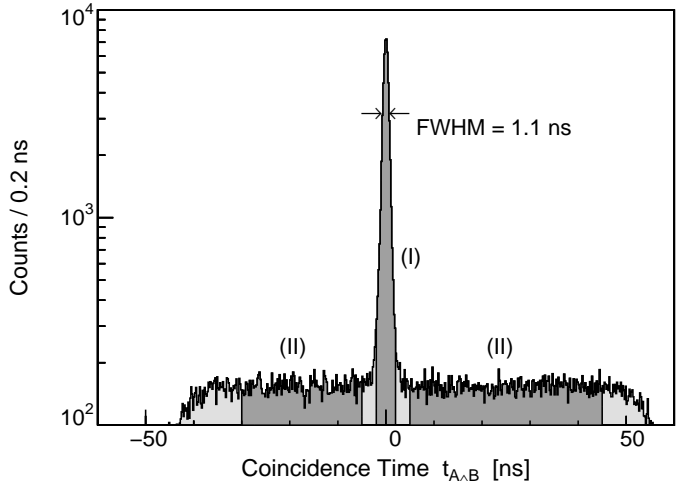}
\caption{Coincidence time distribution (logarithmic scale). Region (I)
  was used as true coincidences, while region (II) was used for the
  estimation of the background contribution by random coincidences.}
\label{figtiming}
\end{figure}

Figure\,\ref{figtiming} shows the distribution of the coincidence time
$|t_{A\wedge B}|$ between spectrometer A and B after correction for
the path length of $\approx 12$\,m in each spectrometer. A clear
coincidence peak with a width of 1.1\,ns FWHM is visible. In addition,
a background of random coincidences of a few percent is visible. In
the analysis, a cut of $|t_{A\wedge B}| < 2$\,ns was used as true
coincidences, while cuts on the side bands of $-30\,\mathrm{ns} <
t_{A\wedge B} < -5\,\mathrm{ns}$ and $5\,\mathrm{ns} < t_{A\wedge B} <
45\,\mathrm{ns}$ were used to determine the contribution of random
coincidences.

The single pion production process was identified by the missing mass
of the pion via the four-momentum balance 
\[
    m_{\mathrm{miss}}^2=\left(e_{\mathrm{in}}+p_{\mathrm{in}}
      -e_{\mathrm{out}}-p_{\mathrm{out}}\right)^2.
\]

\begin{figure}
\includegraphics[width=\columnwidth]{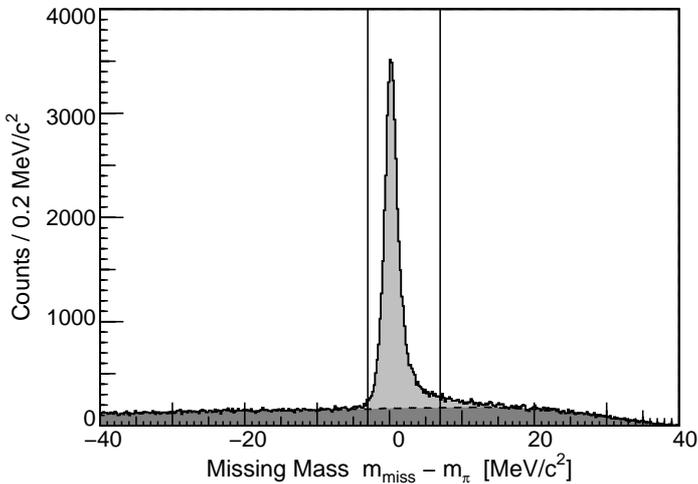}
\caption{Missing mass distribution (for setup
  $Q^2=0.15\,\mathrm{GeV}^2/c^2$). The light shaded area was used as
  pion production yield after subtraction of the background (dark
  shaded area) by random coincidences. The vertical lines show the
  additional cut in missing mass.}
\label{figmissingmass}
\end{figure}

Figure~\ref{figmissingmass} shows the missing mass distribution of one
setup with all events below the timing peak (light shaded area) and
the events of the side band, scaled by the width of the timing window
(dark shaded area). After subtraction of the random coincidences, a
background free missing mass peak remains. A cut $-3\,\mathrm{MeV}/c^2
< m_{\mathrm{miss}}-m_{\pi}<7\,\mathrm{MeV}/c^2$ was used for the identified
production events.

The resulting background subtracted events were histogrammed with
$\Delta W = 1\,$MeV energy bins and eight bins each in the CMS angles
$\cos\theta$ and $\phi$. These bins were compared to a detailed
simulation of the experiment, including the dead-time corrected
luminosity and the cross section of the MAID model \cite{DHKT99} as
generator input. All resolution and line-shape effects, as well as
standard radiative corrections \cite{MoTsai69} were included.

To take the small variation of the cross section within each bin into
account, their relative variations were estimated using models. This
minimized the effects of non-uniform distribution within the bins and
the extracted cross sections can be accurately compared directly to
theoretical calculations at the central value of each bin.

For each bin in $\theta$ and $\Delta W$ a fit of the form \[A+B\,\cos
\phi + \cos 2\phi\] was performed to extract $\sigma_0$ and
$\sigma_{\mathrm{LT}}$ (see eqn.~\ref{equ:cross} for cross section
structure). Since $\sigma_{\mathrm{TT}}$ is small compared to the error bar, it
cannot be extracted from the data and the MAID model value was used as
a constraint for the fit.

\begin{figure}
\centerline{\includegraphics[width=\columnwidth]{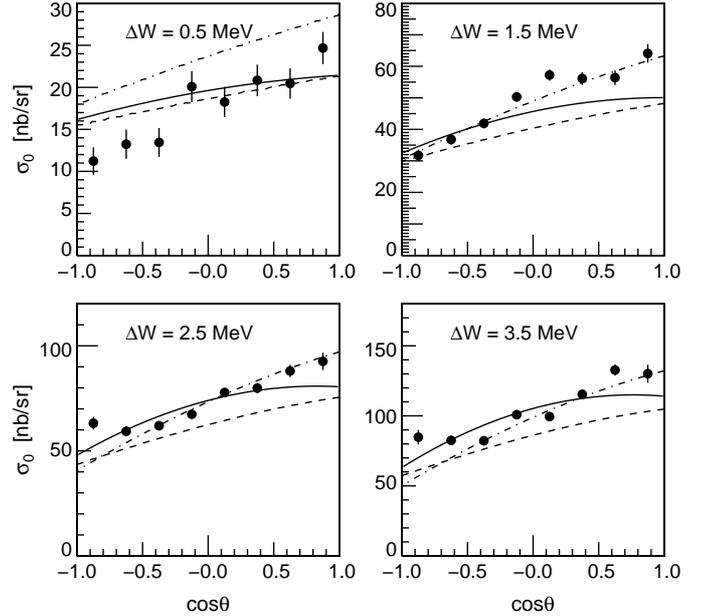}}
\caption{Cross section $\sigma_0$ for
  $Q^2=0.05\,\mathrm{GeV}^2/c^2$. Only statistical errors are
  shown. The solid line is the prediction in HBChPT \cite{BKM96a}, the
  dashed line is the MAID model \cite{DHKT99}, and the dashed-dotted
  line is the DMT model \cite{PhysRevC.64.032201}.}
\label{fig050}
\end{figure}

\begin{figure}
\centerline{\includegraphics[width=\columnwidth]{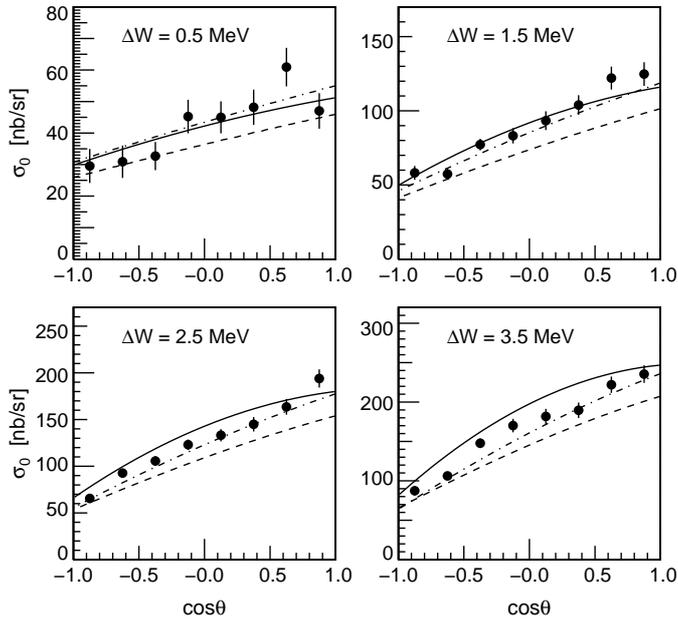}}
\caption{Cross section $\sigma_0$ for $Q^2=0.10\,\mathrm{GeV}^2/c^2$
  (as Fig.~\ref{fig050})
.}
\label{fig100}
\end{figure}

\begin{figure}
\centerline{\includegraphics[width=\columnwidth]{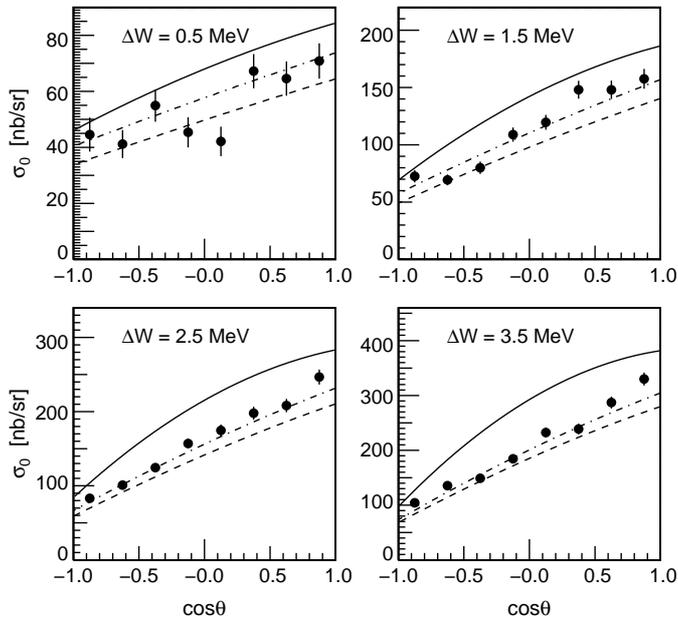}}
\caption{Cross section $\sigma_0$ for $Q^2=0.15\,\mathrm{GeV}^2/c^2$
  (Notation as in Fig.~\ref{fig050}).}
\label{fig150}
\end{figure}

\section{Systematic Errors}

Since consistency was the problem of the previously existing data
sets, special care was taken to minimize and control the systematic
errors.

The detection efficiency of a modern high resolution spectrometer is
above 95\%, thus the efficiency corrections are small and introduce
only systematic errors below the percent level. The beam current was
measured with a fluxgate magnetometer (F\"orster probe) in the
acceleration path of the last microtron stage, i.e.~as the current sum
of 90 staggered turns. By this enhancement of the sensitivity, the
current was measured on the 1\% level. The effective target length
introduced the largest error to the overall normalization of the
cross section by roughly 3\%. In addition to monitoring all these
parameters, the normalization was calibrated for each setup by an
additional measurement of the elastic scattering from the proton using
a standard form factor parametrization \cite{Mergell:1995bf}. These
calibration measurements agreed within 2\% with the calculations.

A major source of systematic errors for threshold measurements is
introduced by the calibration of the momentum measurement of the
electron detection. In the chosen setups, a calibration error of 0.1\%
can cause a cross section error of up to 14\% in the lowest bin by
shifting events below the production threshold. Therefore the momentum
detection of the electron arm was calibrated in situ by a measurement
of the electron momentum in the elastic peak. Since the momentum in
the elastic peak is determined only by the scattering angle, the
absolute momentum could be calibrated to $\pm 50\,$keV central
value. This corresponds to a systematic error of 5\% on the
normalization in the lowest bin. An uncertainty of the incident beam
energy had not to be considered, since a deviation of this energy
would affect the elastic line of the calibration measurement by the
same amount, resulting in no additional uncertainty in the cross
section determination and only a negligible error in the
reconstruction of the four-momentum transfer.

The detection of the electron angle and the momentum and angle of the
proton was calibrated by considering the recoil proton with a narrow
cut on $\Delta W = 2\,\mathrm{MeV}$. As can be seen in
Fig.~\ref{figacceptance}, these events form a ring in the acceptance
of spectrometer A (after correction for different electron scattering
angles) and all misalignments would show up as a shift of this ring.

The estimated systematic errors add up to 10\% for the lowest bin, 5\%
for $\Delta W=1.5\,\mathrm{MeV}$, and $3\%$ for the two highest bins.

\begin{figure}
\centerline{\includegraphics[width=\columnwidth]{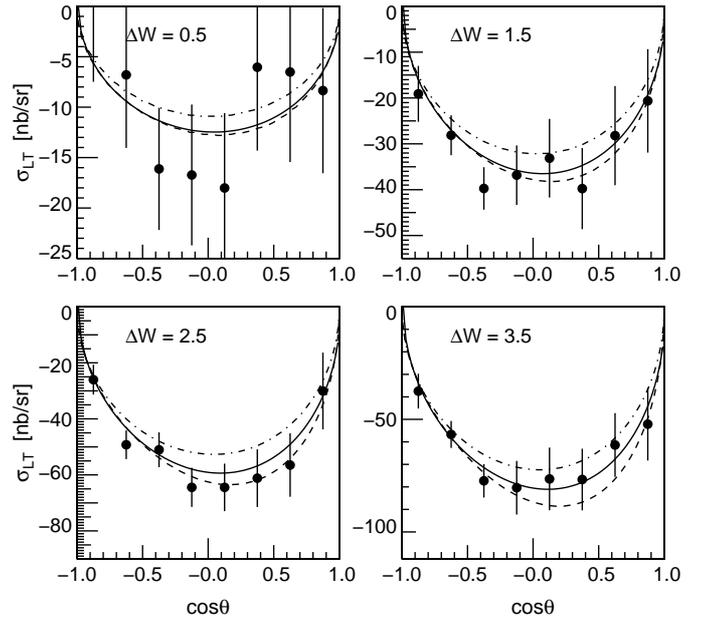}}
\caption{Interference cross section $\sigma_{\mathrm{LT}}$ for
  $Q^2=0.10\,\mathrm{GeV}^2/c^2$ (Notation as in Fig.~\ref{fig050}). }
\label{figLT}
\end{figure}

\section{Results and Discussion}

Figures~\ref{fig050}, \ref{fig100}, and \ref{fig150} show the cross
section $\sigma_0(\theta)$ for the first four energy bins above the
threshold. For comparison, three calculations are included. The dashed
line shows the phenomenological isobar model MAID \cite{DHKT99}. This
model fits basically all existing photo- and electro-production data
and is therefore dominated in the threshold region by the amount of
existing photo-production data. The overall good agreement of the data
with this model can be interpreted as consistency with the world data
set. Even better agreement with the data is shown by the dynamical
model DMT \cite{PhysRevC.64.032201}, which is compared to the MAID model
expected to be superior in the threshold region since it is based on a
chiral Lagrangian.

\begin{figure}
\centerline{\includegraphics[width=\columnwidth]{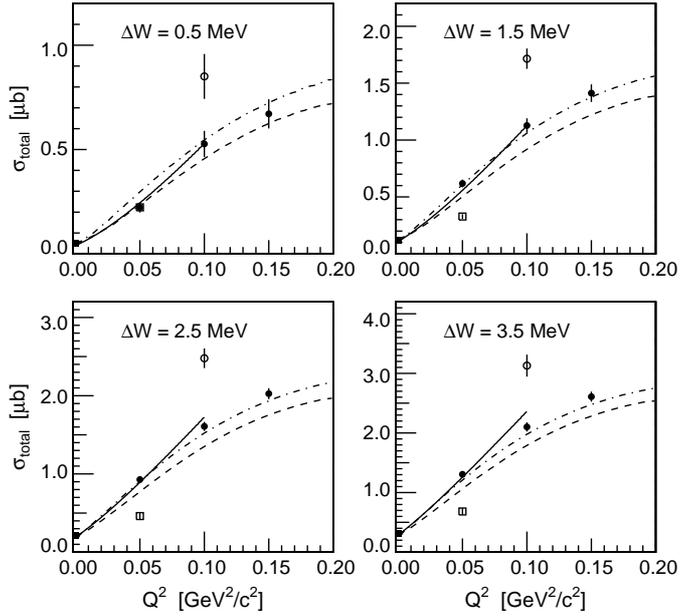}}
\caption{Total cross section $\sigma_{\mathrm{total}}$ versus $Q^2$, calculated
  from a fit of the form $\sigma_0(\theta)=A+B\cos\theta+C\cos^2\theta
  \Rightarrow \sigma_{\mathrm{total}} =4\pi(A+C/3)$. The error bars are the
  quadratic sum of the statistical error and the systematic error. The
  solid line is the prediction in HBChPT \cite{BKM96a}, the dashed
  line is the MAID model \cite{DHKT99}, and the dashed-dotted line is
  the DMT model \cite{PhysRevC.64.032201}. The lines are calculated with
  constant $E_{\mathrm{beam}}=855\,\mathrm{MeV}$, \textit{i.e.}  variable
  photon polarization $\epsilon$. In addition the data points from
  Refs. \cite{Schmidt:2001vg} (photon point), \cite{Merkel:2001qg})
  (open box), and \cite{Distler:1998ae} (open circle) are drawn.}
\label{figq2dep}
\end{figure}

For completeness, the calculation in Heavy Baryon Chiral Dynamics of
Ref.~\cite{BKM96a} is included. This curve has to be interpreted with
care, since the calculation was fitted to the data of
Refs. \cite{Distler:1998ae, vandenBrink:1997cs}, which show serious
consistency problems. These consistency problems are illustrated in
Fig.~\ref{figq2dep}, which shows the total cross section of this
measurement in combination with the photon point \cite{Schmidt:2001vg}
and previous electroproduction data points at
$Q^2=0.05\,\mathrm{GeV}^2/c^2$ (Ref. \cite{Merkel:2001qg}) and
$Q^2=0.10\,\mathrm{GeV}^2/c^2$ (Ref. \cite{Distler:1998ae}, consistent
with Ref. \cite{vandenBrink:1997cs}). The latter was included in the
fit of HBChPT and causes probably the unnatural slope of this
calculation. This is the reason that the results of this calculation
are not shown for $Q^2 > 0.10\,\mathrm{GeV}^2/c^2$ in
Fig.~\ref{figq2dep}. A refit of the empirical low energy constants
with this new data set is needed before any conclusions about its
accuracy can be made.

Figure~\ref{figLT} shows $\sigma_{\mathrm{LT}}$ for the intermediate
$Q^2$ value (figures of the other two settings and all data as table
are available in the online version of this article). Within the error
bar, all three calculations are consistent with the extracted
interference cross section.

In summary, the data sets of Refs. \cite{vandenBrink:1997cs,
  Distler:1998ae, Merkel:2001qg} seem to have  normalization
problems beyond the size of the quoted systematic errors. A reanalysis
of part of this data at $Q^2=0.05\,\mathrm{GeV}^2/c^2$ did not reveal
any obvious mistakes, however, the extensive additional calibration
measurements of the new data set were not done at that time. As
discussed above, e.g.~a plausible small calibration error in the
electron momentum can cause already large normalization errors in the
resulting cross section, which was underestimated at that time. We
recommend to use the new data for future fits of the total cross
section and to use the old data sets only with a free normalization
parameter. The old data sets included a full Rosenbluth separation and
an extraction of $\sigma_{\mathrm{TT}}$ and $\sigma_{\mathrm{LT}}$.
The relative size of these cross sections might not be affected by an
overall normalization problem. The data of Ref. \cite{Weis:2007kf}
with emphasis on higher energies are consistent with the new
measurement.

\section{Summary}

In this letter a new consistent data set on neutral pion
electro-production in the threshold region is presented. The long
standing problem of the normalization of the existing threshold data
sets has been clarified. The new data are described best by the
dynamical model DMT \cite{PhysRevC.64.032201}. However, with a refit of the
low energy constants in HBChPT \cite{BKM96a}, this theory will probably
also be able to describe the data, although the $Q^2$ range of
validity remains still an open question.

This work was supported by the German Research Foundation with the
Collaborative Research Center 443 and by the Humboldt Foundation.

\bibliographystyle{model1-num-names}
\bibliography{paper}

\begin{thebibliography}{23}
\expandafter\ifx\csname natexlab\endcsname\relax\def\natexlab#1{#1}\fi
\providecommand{\bibinfo}[2]{#2}
\ifx\xfnm\relax \def\xfnm[#1]{\unskip,\space#1}\fi
\bibitem[{Donoghue et~al.(1992)Donoghue, Golowich, and Holstein}]{book}
\bibinfo{author}{J.~F. Donoghue}, \bibinfo{author}{E.~Golowich},
  \bibinfo{author}{B.~R. Holstein},
\newblock \bibinfo{journal}{Camb. Monogr. Part. Phys. Nucl. Phys. Cosmol.}
  \bibinfo{volume}{2} (\bibinfo{year}{1992}) \bibinfo{pages}{1--540}.
\bibitem[{Weinberg(1979)}]{Weinberg:1978kz}
\bibinfo{author}{S.~Weinberg},
\newblock \bibinfo{journal}{Physica} \bibinfo{volume}{A96}
  (\bibinfo{year}{1979}) \bibinfo{pages}{327--340}.
\bibitem[{Gasser and Leutwyler(1984)}]{Gasser:1983yg}
\bibinfo{author}{J.~Gasser}, \bibinfo{author}{H.~Leutwyler},
\newblock \bibinfo{journal}{Ann. Phys.} \bibinfo{volume}{158}
  (\bibinfo{year}{1984}) \bibinfo{pages}{142--210}.
\bibitem[{Gasser and Leutwyler(1983)}]{Gasser:1983kx}
\bibinfo{author}{J.~Gasser}, \bibinfo{author}{H.~Leutwyler},
\newblock \bibinfo{journal}{Phys. Lett.} \bibinfo{volume}{B125}
  (\bibinfo{year}{1983}) \bibinfo{pages}{325--329}.
\bibitem[{Bernard and {Mei\ss{}ner}(2007)}]{Bernard:2006gx}
\bibinfo{author}{V.~Bernard}, \bibinfo{author}{U.-G. {Mei\ss{}ner}},
\newblock \bibinfo{journal}{Ann. Rev. Nucl. Part. Sci.} \bibinfo{volume}{57}
  (\bibinfo{year}{2007}) \bibinfo{pages}{33--66}.
\bibitem[{Bernard(2008)}]{Bernard:2007zu}
\bibinfo{author}{V.~Bernard},
\newblock \bibinfo{journal}{Prog. Part. Nucl. Phys.} \bibinfo{volume}{60}
  (\bibinfo{year}{2008}) \bibinfo{pages}{82--160}.
\bibitem[{Bernstein et~al.(2009)Bernstein, Ahmed, Stave, Wu, and
  Weller}]{Bernstein:2009dc}
\bibinfo{author}{A.~M. Bernstein}, \bibinfo{author}{M.~W. Ahmed},
  \bibinfo{author}{S.~Stave}, \bibinfo{author}{Y.~K. Wu},
  \bibinfo{author}{H.~R. Weller},
\newblock \bibinfo{journal}{Ann. Rev. Nucl. Part. Sci.} \bibinfo{volume}{59}
  (\bibinfo{year}{2009}) \bibinfo{pages}{115--144}.
\bibitem[{Schmidt et~al.(2001)}]{Schmidt:2001vg}
\bibinfo{author}{A.~Schmidt}, et~al.,
\newblock \bibinfo{journal}{Phys. Rev. Lett.} \bibinfo{volume}{87}
  (\bibinfo{year}{2001}) \bibinfo{pages}{232501}.
\bibitem[{Bernard et~al.(1996)Bernard, Kaiser, and
  {Mei\ss{}ner}}]{Bernard:1994gm}
\bibinfo{author}{V.~Bernard}, \bibinfo{author}{N.~Kaiser},
  \bibinfo{author}{U.-G. {Mei\ss{}ner}},
\newblock \bibinfo{journal}{Z. Phys.} \bibinfo{volume}{C70}
  (\bibinfo{year}{1996}) \bibinfo{pages}{483--497}.
\bibitem[{Hornidge and Bernstein(2011)}]{hornidge}
\bibinfo{author}{D.~Hornidge}, \bibinfo{author}{A.~M. Bernstein}
  (\bibinfo{year}{2011}). \bibinfo{note}{{arXiv:1108.6029v1 [nucl-ex]}}.
\bibitem[{Welch et~al.(1992)}]{Welch:1992ex}
\bibinfo{author}{T.~P. Welch}, et~al.,
\newblock \bibinfo{journal}{Phys. Rev. Lett.} \bibinfo{volume}{69}
  (\bibinfo{year}{1992}) \bibinfo{pages}{2761--2764}.
\bibitem[{van~den Brink et~al.(1997)}]{vandenBrink:1997cs}
\bibinfo{author}{H.~B. van~den Brink}, et~al.,
\newblock \bibinfo{journal}{Nucl. Phys.} \bibinfo{volume}{A612}
  (\bibinfo{year}{1997}) \bibinfo{pages}{391--417}.
\bibitem[{Distler et~al.(1998)}]{Distler:1998ae}
\bibinfo{author}{M.~O. Distler}, et~al.,
\newblock \bibinfo{journal}{Phys. Rev. Lett.} \bibinfo{volume}{80}
  (\bibinfo{year}{1998}) \bibinfo{pages}{2294--2297}.
\bibitem[{Bernard et~al.(1996)Bernard, Kaiser, and Mei\ss{}ner}]{BKM96a}
\bibinfo{author}{V.~Bernard}, \bibinfo{author}{N.~Kaiser},
  \bibinfo{author}{U.-G. Mei\ss{}ner},
\newblock \bibinfo{journal}{Nucl. Phys.} \bibinfo{volume}{A607}
  (\bibinfo{year}{1996}) \bibinfo{pages}{379--401}. \bibinfo{note}{Erratum A633
  (1998) 695-697}.
\bibitem[{Merkel et~al.(2002)}]{Merkel:2001qg}
\bibinfo{author}{H.~Merkel}, et~al.,
\newblock \bibinfo{journal}{Phys. Rev. Lett.} \bibinfo{volume}{88}
  (\bibinfo{year}{2002}) \bibinfo{pages}{012301}.
\bibitem[{Weis et~al.(2008)}]{Weis:2007kf}
\bibinfo{author}{M.~Weis}, et~al.,
\newblock \bibinfo{journal}{Eur. Phys. J.} \bibinfo{volume}{A38}
  (\bibinfo{year}{2008}) \bibinfo{pages}{27--33}.
\bibitem[{Drechsel et~al.(1999)Drechsel, Hanstein, Kamalov, and
  Tiator}]{DHKT99}
\bibinfo{author}{D.~Drechsel}, \bibinfo{author}{O.~Hanstein},
  \bibinfo{author}{S.~S. Kamalov}, \bibinfo{author}{L.~Tiator},
\newblock \bibinfo{journal}{Nucl. Phys.} \bibinfo{volume}{A645}
  (\bibinfo{year}{1999}) \bibinfo{pages}{145--174}.
\bibitem[{Kamalov and Nan~Yang(1999)}]{PhysRevLett.83.4494}
\bibinfo{author}{S.~S. Kamalov}, \bibinfo{author}{S.~Nan~Yang},
\newblock \bibinfo{journal}{Phys. Rev. Lett.} \bibinfo{volume}{83}
  (\bibinfo{year}{1999}) \bibinfo{pages}{4494--4497}.
\bibitem[{Kamalov et~al.(2001)Kamalov, Nan~Yang, Drechsel, Hanstein, and
  Tiator}]{PhysRevC.64.032201}
\bibinfo{author}{S.~S. Kamalov}, \bibinfo{author}{S.~Nan~Yang},
  \bibinfo{author}{D.~Drechsel}, \bibinfo{author}{O.~Hanstein},
  \bibinfo{author}{L.~Tiator},
\newblock \bibinfo{journal}{Phys. Rev.} \bibinfo{volume}{C64}
  (\bibinfo{year}{2001}) \bibinfo{pages}{032201}.
\bibitem[{Drechsel and Tiator(1992)}]{DT92}
\bibinfo{author}{D.~Drechsel}, \bibinfo{author}{L.~Tiator},
\newblock \bibinfo{journal}{J. Phys.} \bibinfo{volume}{G18}
  (\bibinfo{year}{1992}) \bibinfo{pages}{449--497}.
\bibitem[{Blomqvist et~al.(1998)}]{Blo98}
\bibinfo{author}{K.~I. Blomqvist}, et~al.,
\newblock \bibinfo{journal}{Nucl. Instrum. Meth.} \bibinfo{volume}{A403}
  (\bibinfo{year}{1998}) \bibinfo{pages}{263--301}.
\bibitem[{Mo and Tsai(1969)}]{MoTsai69}
\bibinfo{author}{L.~W. Mo}, \bibinfo{author}{Y.-S. Tsai},
\newblock \bibinfo{journal}{Rev. Mod. Phys.} \bibinfo{volume}{41}
  (\bibinfo{year}{1969}) \bibinfo{pages}{205--235}.
\bibitem[{Mergell et~al.(1996)Mergell, {Mei\ss{}ner}, and
  Drechsel}]{Mergell:1995bf}
\bibinfo{author}{P.~Mergell}, \bibinfo{author}{U.-G. {Mei\ss{}ner}},
  \bibinfo{author}{D.~Drechsel},
\newblock \bibinfo{journal}{Nucl. Phys.} \bibinfo{volume}{A596}
  (\bibinfo{year}{1996}) \bibinfo{pages}{367--396}.

\end{thebibliography}
\end{document}